\documentclass[reprint]{revtex4-1}

\usepackage[utf8]{inputenc}
\usepackage{amsmath}
\usepackage{amssymb}
\usepackage{booktabs}
\usepackage{graphicx}
\usepackage{listings}
\usepackage{units}
\usepackage{wasysym}

\graphicspath{{./}{img/}}

\begin{document}

\title{A Parallelized Vlasov-Fokker-Planck-Solver for Desktop PCs}
\date{\today}

\author{Patrik Schönfeldt}
\email{patrik.schoenfeldt@kit.edu}
\author{Miriam Brosi}
\author{Markus Schwarz}
\author{Johannes L. Steinmann}
\author{Anke-Susanne Müller}
\affiliation{Karlsruhe Institute of Technology,
Kaiserstraße 12, 76131 Karlsruhe, Germany}

\begin{abstract}
The numerical solution of the Vlasov-Fokker-Planck equation is
a well established method to simulate the dynamics, including
the self-interaction with its own wake field, of an electron bunch
in a storage ring.
In this paper we present \emph{Inovesa}, a modularly extensible
program that uses OpenCL to massively parallelize the computation.
It allows a standard desktop PC to work with appropriate accuracy
and yield reliable results within minutes.
We provide numerical stability-studies over a wide parameter range
and compare our numerical findings to known results.
Simulation results for the case of coherent synchrotron radiation
will be compared to measurements that probe
the effects of the micro-bunching instability
occurring in the short bunch operation at ANKA.
It will be shown that the impedance model based
on the shielding effect of two parallel plates
can not only describe the instability threshold,
but also the presence of multiple regimes that
show differences in the emission of coherent synchrotron radiation.
\end{abstract}

\maketitle

\section{Introduction}

At synchrotron light sources electron bunches of a length of a few millimeters
are used to produce coherent synchrotron radiation (CSR) in the terahertz (THz)
frequency range.
Due to the coherent emission, the intensity scales with the number of
emitting particles squared, instead of linearly as for incoherent emission.
In storage rings the spatial compression is achieved by using
magnet optics with a small momentum compaction factor \(\alpha_c\).
The compression leads to the micro-bunching instability.
On the one hand, this instability limits the electron bunch charge
that can be used in stable operation;
on the other hand the emerging substructures emit coherent radiation
also in wavelength smaller than the electron bunch length.

First observations of micro-bunching in a storage ring
as well as of the increase of
coherent emission in the spectral range of interest were made at
the NSLS VUV ring~\cite{carr1999,carr-2001}.
It then has been studied both experimentally,
to map out the parameters governing the bursting behavior
(e.g. at ANKA~\cite{anke2010}, BESSY II~\cite{bakr2003},   
DIAMOND~\cite{diamond_bursting_2012}, MLS~\cite{mls_ipac10},
and SOLEIL~\cite{soleil_2012}),
and theoretically to predict
thresholds~\cite{venturini2002, Stupakov2002, Bane_cai_stupakov2010, cai2011ipac}
and to simulate the dynamics.
To simulate the dynamics, it is possible to solve
the Vlasov-Fokker-Planck equation for the longitudinal phase space
density~\cite{PhysRevSTAB.8.014202,PhysRevSTAB.17.010701},
or, using supercomputers, to do particle tracking
with one million macro particles or more~\cite{0295-5075-98-4-40006}.

Recent advances in detector development and readout electronics facilitate
fast mapping of the micro-bunching instability~\cite{2016arXiv160500536B}
over a wide range of physical parameters.
To cover these settings in simulation as well, it calls
for an ultra-fast simulation technique.
Also, the simulation tool should be designed such that
the influences of both the simulated physics and of numerical effects
can be studied and separated.
As we are interested in simulating an instability,
in particular sources of numerical instabilities should be ruled out.
In this paper we present \emph{Inovesa}
(\emph{Inovesa Numerical Optimized Vlasov-Equation Solver Application},
available at \cite{github-inovesa}),
a Vlasov-Fokker-Planck solver that runs on standard desktop PCs
and yields robust results within minutes.

Section~\ref{sec: physics} summarizes the theoretical description of the problem.
The actual implementation is described in section~\ref{sec: implementation},
while section~\ref{sec: results} presents numerical studies that
show the robustness of the implementation
and compares results of simulation and measurements.

\section{Longitudinal Phase-Space Dynamics}
\label{sec: physics}

\subsection{Vlasov-Fokker-Planck equation}

The phenomenon of micro-bunching happens in the longitudinal phase space,
which is spanned by the position \(z\) relative to the synchronous particle
and the energy \(E\). 
Taking the particle density \(\psi(z,E,t)\) of electrons in a storage ring
to be a smooth function, its evolution with time \(t\) can be described by the
Vlasov-Fokker-Planck equation (VFPE).
Following the notation of~\cite{Bane_cai_stupakov2010} it reads
\begin{equation}
	\frac{\partial \psi}{\partial \theta}
	+ \frac{\partial H}{\partial p} \frac{\partial \psi}{\partial q}
	- \frac{\partial H}{\partial q} \frac{\partial \psi}{\partial p}
	= \beta \frac{\partial}{\partial p}\left(p\psi + \frac{\partial \psi}{\partial p}\right),
	\label{eq: VFP}
\end{equation}
with the time given in multiples of synchrotron periods \(\theta = f_s t\),
the normalized coordinates \(q={z}/{\sigma_{z,0}}\), and
\(p={(E-E_0)}/{\sigma_{\delta,0}}\), the Hamiltonian \(H\),
the reference particle's energy \(E_0\),
and \(\beta = 1/(f_s \tau_d)\), where \(\tau_d\)
is the longitudinal damping time.
The quantities \(\sigma_{\delta,0}\) and \(\sigma_{z,0}\)
describe, respectively, energy spread and bunch length
in the equilibrium state
that exists for small bunch charges.

To solve this partial differential equation
there exists a widely used formalism by Warnock and Ellison~\cite{slac-pub-8404}.
It uses a grid to discretize \(\psi(q,p)\) and 
assumes that the collective force due to self-interaction with the bunch's own
coherent synchrotron radiation is constant for small time steps.
The perturbation due to the collective effects is described
as a perturbation to the Hamiltonian
\begin{align}
	H(q,p,t)
		&= \underbrace{H_e(q,p,t)}_{\mathrm{external\;fields}}
		 + \underbrace{H_c(q,t)}_{\mathrm{collective\;effects}}\nonumber\\
		&= \frac{1}{2}\left(q^2+p^2\right)
			+ Q_c\times V_c(Z_c,q,t),
	\label{eq: perturbation}
\end{align}
where \(Q_c\) is the charge involved in the perturbation, and
\(V_c\) is the potential due to the collective effect, which 
can be expressed in terms of an impedance \(Z_c\).

It is then possible to use the homogeneous solution,
which in the unperturbed case is represented by a rotation in phase space,
and add the influence of diffusion
and damping as a particular solution.
To model the perturbation, the influence of \(V_c\) is implemented as a `kick'
along the energy axis.

\subsection{Micro-Bunching Instability}

To calculate the effect of the perturbation term
introduced in Eq.~\ref{eq: perturbation},
one needs the electric field \(E(q,s)\)
at the longitudinal position \(s\).
It is convenient to express it via a wake potential
\begin{align}
	V(q)	&= \int E(q,s)\, \mathrm{d}s,
\end{align}
which directly gives the energy difference for the electrons (in \(\unit{eV}\)).
The wake potential can be obtained from the wake function \(W(q)\),
which describes the field produced by one single particle.
The wake potential \(V(q)\) then is obtained
by convolving it with the charge density \(\varrho\)~\cite{slac-pub-8404}
\begin{equation}
	V(q) = \int_{-\infty}^{\infty} W(q-q') \varrho(q')\,\mathrm{d}q'.
\end{equation}

As in frequency space closed and smooth expressions exist for many
commonly used impedances, we decided to work in frequency space.
Then the wake potential can be deduced directly from the impedance \(Z(k)\)
in every time step using
\begin{equation}
	V(q) = \int_{-\infty}^{\infty} Z(k) \tilde{\varrho}(k) e^{ikq}\,\mathrm{d}k,
	\label{eq: wakepotential}
\end{equation}
where \(\tilde{\varrho}(k)\) is the Fourier transform of the bunch profile.

This method allows to implement different impedance models
for \emph{Inovesa} in just a few lines of code.
As we are mostly interested in CSR-driven dynamics,
we currently implemented two cases.
The first, the free space CSR impedance,
describes the effect of coherent synchrotron radiation of
particles traveling on a curved path in vacuum.
A good approximation for the CSR impedance
in a perfect circle is~\cite{murphy-wakefield}
\begin{equation}
	Z(n) \approx Z_0 \frac{\Gamma(2/3)}{3^{1/3}}
				\left(\frac{\sqrt{3}}{2}+\frac{i}{2}\right) n^{1/3},
\end{equation}
where \(Z_0\) is the vacuum impedance,
and \(n=f/f_{rev}\) is frequency expressed
in multiples of the revolution frequency \(f_{rev}\).

The second implemented impedance approximates the shielding effect
of the beam pipe by two parallel plates with distance \(g\)~\cite{murphy-wakefield}.
It can be approximated
with the Airy functions Ai and Bi~\cite{IPAC11-FRXAA01}:
\begin{align}
	Z(n,R/g) 	&\approx \frac{4 \pi^2\, 2^{1/3}}{\epsilon_0 c}
					\left(\frac{R}{g}\right) n^{-1/3}\nonumber\\
				&\times\sum\limits_{p} \mathrm{Ai}'(u_p)\mathrm{Ci}'(u_p)
				+u_p\;\mathrm{Ai}(u_p)\mathrm{Ci}(u_p),
\end{align}
where the prime marks the first derivative with respect to the argument,
\(R\) is the beam path's radius,
\(\mathrm{Ci} := \mathrm{Ai}-j\,\mathrm{Bi}\), and
\begin{equation}
	u_p := \frac{\pi^2 \left(2p+1\right)^2}{2^{2/3}}\left(\frac{R}{g}\right)^2 n^{-4/3}.
\end{equation}
\begin{figure}[bt]
\centering
\includegraphics[width=\linewidth]{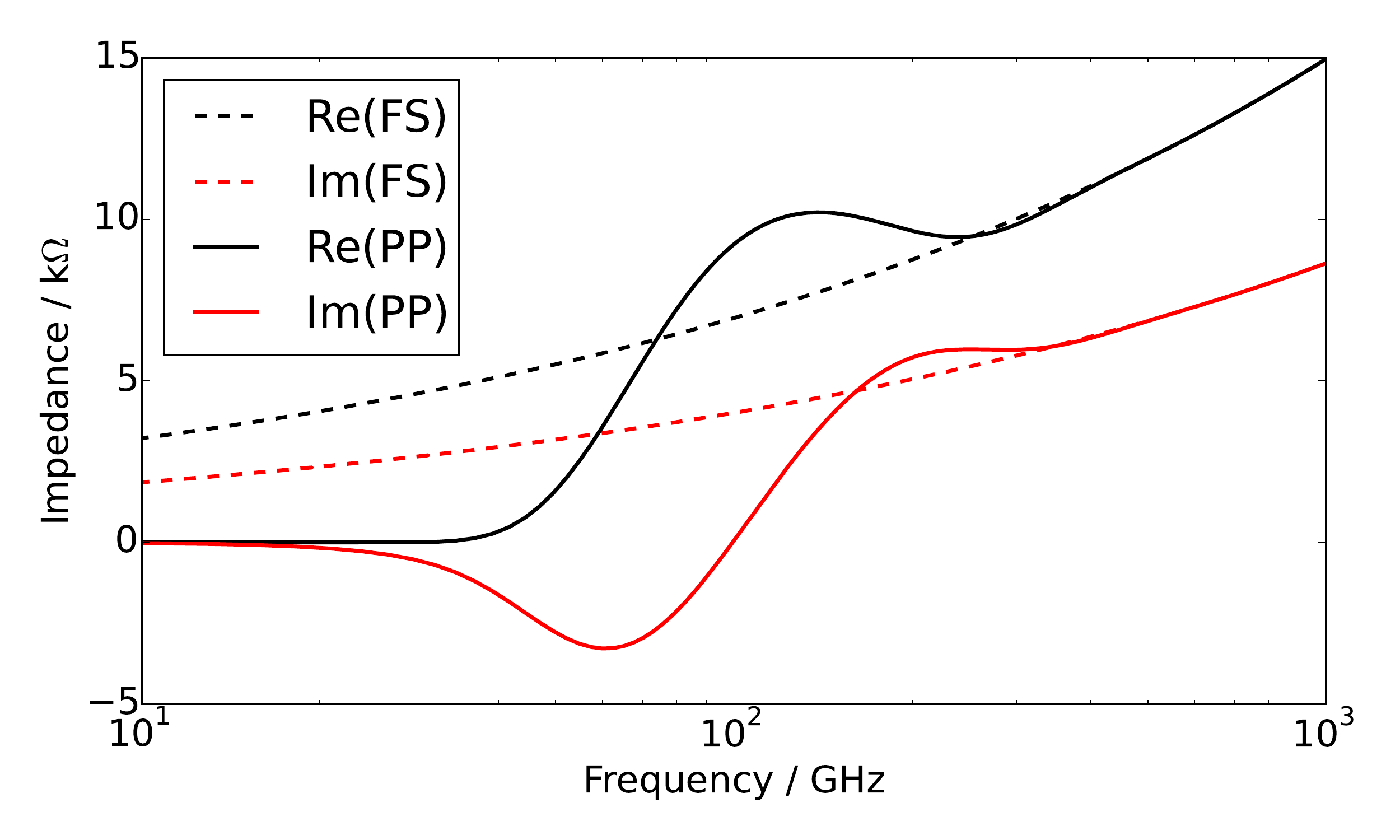}
\caption{
Unshielded (free space) CSR impedance (FS) and CSR impedance shielded by
parallel plates (PP), calculated for the case of an accelerator
with \(f_{rev}=\unit[8.582]{GHz}\), and (for the shielded case)
\(g=\unit[32]{mm}\).
For high frequencies both impedances converge,
as short wavelength are not effected by the shielding.
For \(f \rightarrow \unit[0]{Hz}\) both
impedances approach \(Z=\unit[0]{\Omega}\).
}
\label{fig: impedance}
\end{figure}
This more complex impedance has already proven to describe the
micro-bunching instability threshold within the uncertainty
of the measurements~\cite{2016arXiv160500536B}.
For the limit \(g \rightarrow \infty\) it converges to the free space impedance.
Figure~\ref{fig: impedance} shows these two impedances,
for the case of an accelerator with \(f_{rev}=\unit[8.582]{GHz}\),
and (for the shielded case) \(g=\unit[32]{mm}\).

\section{Implementation}
\label{sec: implementation}

\subsection{Discretization}

Following the approach of Warnock and Ellison~\cite{slac-pub-8404},
the VFPE is discretized on a grid to be solved numerically.
For \emph{Inovesa},
we define a grid starting from a minimum value that can be expressed
for each dimension.
For \(q > q_{\mathrm{min}}\) and \(p > p_{\mathrm{min}}\),
the generalized coordinates \(q, p \in \mathbb{R}\)
become the grid coordinates \(x_r, y_r \in \mathbb{R}_{\ge0}\).
With \(\Delta p\), the granularity of the grid in energy direction,
the transformation between the coordinate systems can be expressed as
\begin{equation}
	y_r(p) = (p-p_{\mathrm{min}})/\Delta p,
	\label{eq: coordinate trafo}
\end{equation}
and accordingly for \(x_y(q)\). 
Function values are only stored at integer coordinates \(m\),
where \(m\) refers to either \(x\) or \(y\).
When a function value at an arbitrary
non-integer coordinate \(m_r\) is needed, Inovesa approximates it by
interpolation using
\begin{equation}
	f(m_r) \approx \vec{f}_N(m)\cdot\vec{p}_N(\{m\}) =: P_N(m_r)
	\label{eq: interpolation}
\end{equation}
were \(\{m\}=m_r - m\) denotes the fractional part of \(m_r\).
The interpolation multiplies the vector of the function values at the \(N\)
surrounding mesh points
\begin{align}
\vec{f}_N(m) =	(
					&f(m-\lfloor (N-1)/2\rfloor), \dots, f(m),\nonumber \\
					&\dots, f(m+\lceil (N-1)/2\rceil)
				)^T
\end{align}
with a vector containing \(N\) interpolation coefficients
\begin{equation}
	\vec{p}_N(\{m\}) = (l_{0,N}(\{m\}), \dots, l_{N-1,N}(\{m\}))^T,
\end{equation}
where \(l_{\nu,N}(\{m\})\) are the
Lagrange basis polynomials~\cite{numericalrecipes}
\begin{equation}
	l_{\nu,N}(\{m\}) :=	\prod\limits_{k=0\neq \nu}^{(N-1)}
								\frac{\{m\}-m_k}{m_\nu - m_k}.
	\label{eq:lagrange-basis-polynomials}
\end{equation}
We have implemented this for up to \(N = 4\), which results in a cubic
interpolation scheme.
Interpolation using these polynomials sometimes overshoots the values
of the neighboring grid cells.
Section~\ref{ssect: numerical artifacts} will show
that these numerical artifacts can even self-amplify and become a very dominant
feature in the simulation.
To avoid overshooting,
one has to use clamped or saturated interpolation functions.
A simple, clamped version of Eq.~\ref{eq: interpolation} reads
\begin{align}
	\mathrm{clamp}\left(f(m_r)\right) =
	&\max\{
		\min[
			\min(f_N(m),f_N(m+1)),\nonumber\\
			&\vec{f}_N(m)\cdot\vec{p}_N(\{m\})],\nonumber\\
		&\max(f_N(m),f_N(m+1))
	\}.
	\label{eq: clamping}
\end{align}

Keeping in mind that not only \(q,p \in \mathbb{R}\) is discretized
to \(x, y \in \mathbb{N}\)
but also \(\psi(x,y)\) is discretized in the computer's memory,
we analyzed the effect of this second discretization.
To be able to change the accuracy in small steps, we used
a fixed point representation~\cite{fpml-web} where the number of bits for the
fractional part could be chosen freely.
The result of this test will be shown in the convergence studies
(section~\ref{ssec: convergence studies}).

\subsection{Simulation steps}
\label{ssec: simulation steps}

Each time step \(f: \psi_t(q,p) \rightarrow \psi_{t+\Delta t}(q,p)\) is composed
of a number of simulation steps that model rotation, kick, damping and diffusion
\(f = f_{rot} + f_{kick} + f_{damp,diff}\).
\emph{Inovesa} splits the direct implementation
of each of these simulation steps
into two sub-steps.
The information on the actual coordinates (\(q,p\)) is
used by the first half simulation step.
We call its result a `source map' (SM).
Then, in the second half step only the grid coordinates \(x,y\) are used.
Further we define \(z=N_y\times x+y\)
with the number of grid cells in energy direction \(N_y\).
Doing so, the function
\(f: \psi_t(\mathbb{R}^2) \rightarrow \psi_{t+\Delta t}(\mathbb{R}^2)\)
becomes a one dimensional function
\(f_{SM}: \psi_t(\mathbb{R}^1) \rightarrow \psi_{t+\Delta t}(\mathbb{R}^1)\),
which depends only on \(z\).

This method -- by construction --
produces the same results as the single-step implementation.
Practically speaking, the source map expresses the information which
data of the current simulation step contributes
to a grid point for the next simulation step directly
in terms of position in the computer's memory.
For many functions -- such as rotation -- the SM
will look the same for the whole runtime of the program.
For that reason, it only has to be computed once
during a simulation run and can be kept
for multiple usages.

\begin{figure}[b]
\centering
\includegraphics[width=\linewidth]{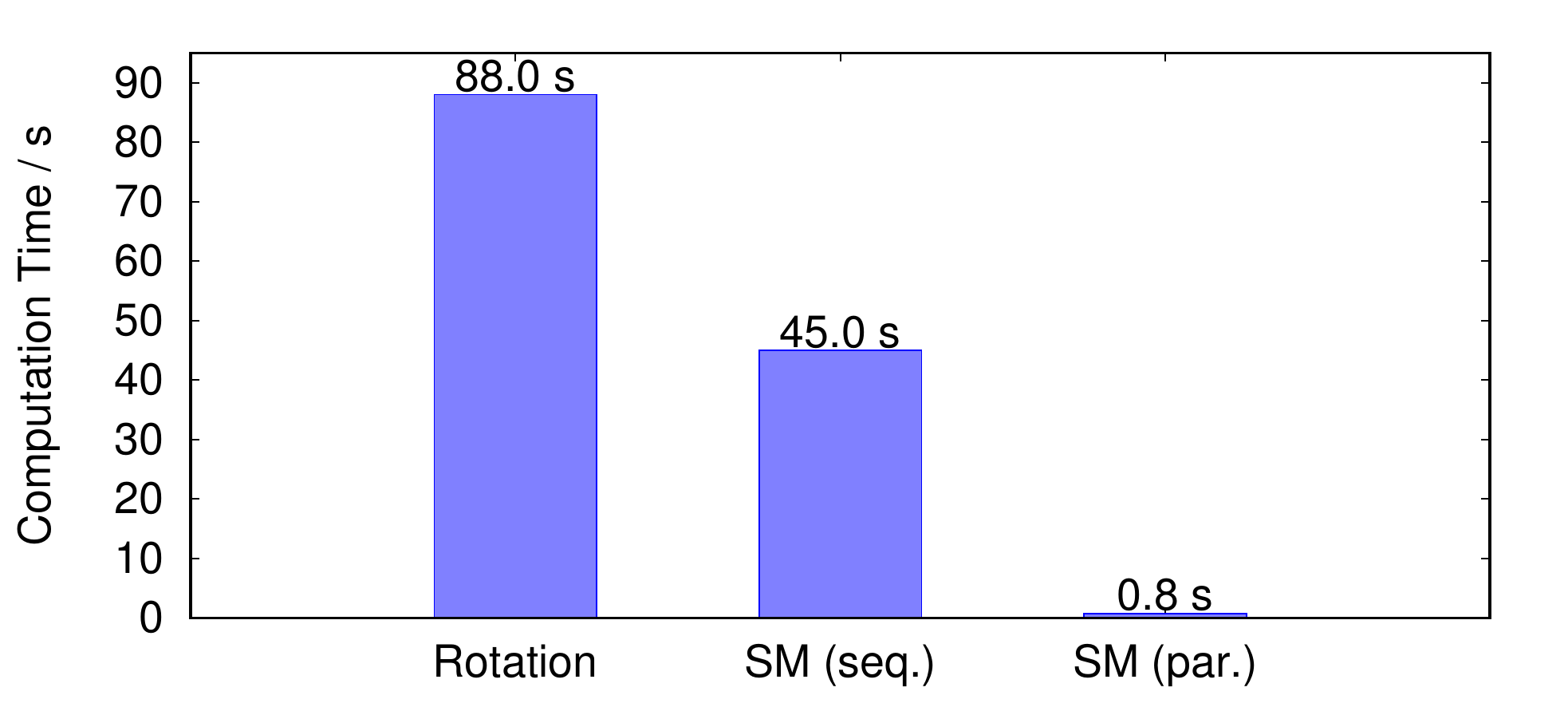}
\caption{Computational time needed by an Intel Core i5 4258U
for 1000 cubically interpolated rotation steps using different implementations.
In this test case, the grid has 512 points per axis,
no optimizations (besides SM) were used.
The first bar represents the computational time
a direct implementation of the rotation takes.
The second and third bar show the time the implementation using
the source map formalism take -- running sequentially or
parallel on the integrated graphics processor.}
\label{fig: sm-performance}
\end{figure}

The source map formalism does not only allow to keep intermediate results,
it also gives a handy interface to implement arbitrary functions
on the phase space.
Furthermore, the reduction of the problem's dimension also leads to a speedup
of the calculations.
Results of a benchmark of the computational performance
for the particular case of rotation are shown in
Fig.~\ref{fig: sm-performance}.
In this case source mapping halves the runtime.
Parallelization using OpenCL~\cite{opencl-web} allows further speedup.
One advantage of OpenCL is
that it can utilize not only multi-core CPUs
but also graphic processors.
In total, a non-optimized program takes days for a typical simulation run.
Using the method described above, \emph{Inovesa} can reduce this to 15 minutes
when running on a customer grade graphics card.

Using the SM formalism, we implemented different versions of
the simulation steps necessary to solve Eq.~\ref{eq: VFP}.
For the rotation, we provide a direct implementation
(as in~\cite{slac-pub-8404}, `standard rotation').
Additionally, we use a symplectic integrator~\cite{wolski-beam_dynamics}
to implement the rotation.
This method offers two advantages:
First, since the method is symplectic, it is automatically area preserving --
also for truncated power series.
Using the map from an infinitesimal rotation given by a direct implementation,
symplecticity is easily lost in the numeric treatment.
Second, the symplectic method provides additional numerical stability
also for the interpolation.
The standard rotation algorithm requires a two-dimensional interpolation,
which involves \(N\times N\) data points,
leading to interpolation coefficients
\(l_{\nu,N}^2 \propto \{x\}^{N-1}\times\{y\}^{N-1} \ll 1\).
The symplectic map, in contrast, splits the rotation into a energy-dependent
drift followed by a location dependent RF kick.
Each requires a one-dimensional interpolation,
involving \(N\) data points.
The resulting coefficients are \(l_{\nu,N}^1 \propto \{m\}^{N-1} \gg l_{\nu,N}^2\),
minimizing the  vulnerability to numerical absorption.
Since a point is rotated by moving on straight lines in perpendicular directions,
in the following the symplectic
approach will be referred to as `Manhattan rotation'.

The damping and diffusion terms (right hand side of Eq.~\ref{eq: VFP})
need numerical differentiation.
We found that the same type of artifacts that we found for the interpolation
(see section~\ref{ssect: numerical artifacts})
can also occur because of the differentiation.
This can be explained by the fact that the algorithm usually used for
numerical differentiation~\cite{numericalrecipes}
is equivalent to differentiating the
quadratic interpolation polynomial \(P_3\) (see Eq.~\ref{eq: interpolation})
\begin{equation}
	\frac{\partial f(x)}{\partial x}\biggr\rvert_{x_0}
	\approx \frac{f(x_0+\Delta x)-f(x_0-1)}{2 \Delta x}
	= \frac{\partial P_3(x)}{\partial x}\biggr\rvert_{x_0},\nonumber
\end{equation}
where the distance between the sampling points
in our case is \(\Delta x = 1\).
As a consequence, we target this by using the cubic interpolation
polynomial \(P_4\), and obtain
\begin{multline}
	\frac{\partial f(x)}{\partial x}\biggr\rvert_{x_0}
	\approx \frac{-2f(x_0-\Delta x)
			-3 f(x_0)}{6\Delta x}\\
	+ \frac{6 f(x_0+\Delta x)
			-f(x_0+2\Delta x)}{6\Delta x}
	= \frac{\partial P_4(x)}{\partial x}\biggr\rvert_{x_0}.
	\label{eq: cubic diff}
\end{multline}
Aside from this improvement, we proceed analogously to~\cite{slac-pub-8404}.

To implement the perturbation via a kick,
we just had to translate the wake potential
(see Eq.~\ref{eq: wakepotential}) to the grid coordinate system using
Eq.~\ref{eq: coordinate trafo}.
Furthermore, our implementation uses the fact that
both \(Z(k)\) and \(\tilde{\varrho}(k)\) are Hermitian.
This means that optimized algorithms like
the ones from Ref.~\cite{fftw-web,clfft-web} only need
explicit function values for \(k\ge 0\)
to perform the inverse Fourier transform.
Using this symmetry also brings an improvement in both speed and memory usage
by roughly a factor of two~\cite{fftw-web}.

\section{Results}
\label{sec: results}

\subsection{Convergence studies}
\label{ssec: convergence studies}

In this section we compare the effect of different numerical settings,
which ideally should not affect the physical result.
We also compare different implementations
of the rotation as described in section~\ref{ssec: simulation steps}.
For the sake of simplicity, we go to the unperturbed case
(meaning \(H_c = 0\) in Eq.~\ref{eq: perturbation}).
So any starting distribution should exponentially converge to
a Gaussian distribution with \(\sigma_q = \sigma_p = 1\).

\begin{figure}[bt]
\centering
\includegraphics[width=\linewidth]{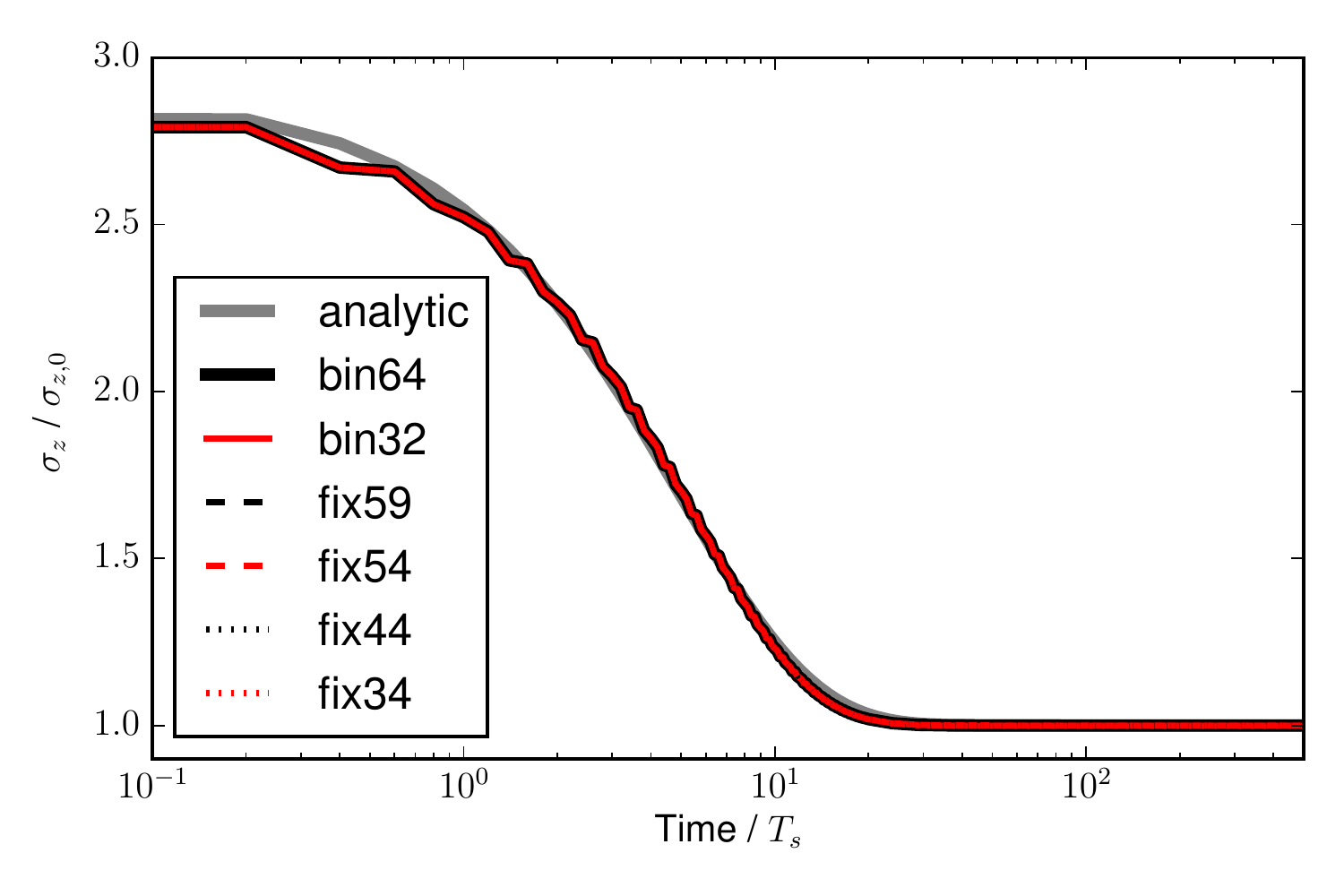}
\caption{
Exponential damping of the bunch length \(\sigma_z\)
as a function of time for different data types.
There is an initial jitter due to quantization noise in the test pattern
that has been read in from a 16 bit \emph{PNG} file
and due to the fact that the initial distribution is not
fully covered by the grid.
Even with this problematic starting conditions,
after \(T=\unit[60]{T_s}\),
all data types have converged
to a constant value.
Note that all simulation curves are almost perfectly overlapping,
there is no noticeable difference coming from the used data type.
The values the simulations converge to are listed in
Tab.~\ref{tab: datatype offset}.
}
\label{fig: data-type}
\end{figure}

\begin{table}[bt]
\caption{Deviation of the converged simulation results
for the different used data types
shown in Fig.~\ref{fig: data-type} from
the analytic result (\(\unit[1]{\sigma_{z,0}}\)).}
\begin{center}
\begin{tabular}{l | c}
Data Type	& Deviation / \(\sigma_{z,0}\)\\
\hline
bin64 & \(0.991\times10^{-4}\)\\
bin32 & \(0.998\times10^{-4}\)\\
fix59 & \(0.991\times10^{-4}\)\\
fix54 & \(0.991\times10^{-4}\)\\
fix44 & \(0.999\times10^{-4}\)\\
fix34 & \(8.877\times10^{-4}\)
\end{tabular}
\label{tab: datatype offset}
\end{center}
\end{table}

At first, we investigate  the effect on the results
of using different data types.
To do so, we observe the evolution
of a Gaussian distributed charge density
with \(\sigma_q = \sigma_p = 2.8\), when the damping time
is set to five synchrotron periods \(\tau=5\,T_s\).
We start by reading the distribution from a 16-bit grayscale PNG.
This brings initial quantization noise,
but also provides well defined starting conditions for every data type:
Initial rounding errors will be the same in the different runs.
Figure~\ref{fig: data-type} shows the different simulation results.
We find that the results obtained using fixed point representations
with a fractional part of at least 44 bits,
and the tested floating point representations
(\texttt{binary32} and \texttt{binary64}~\cite{IEEE_754-2008},
often referred to as `single precision' and `double precision')
show a common difference from the analytic result of about
\(10^{-4}\).
The relative differences between the data types are
less then \(10^{-6}\) and therefore can be neglected.
As most libraries are developed focusing on floating point data types,
we implemented the calculation of the wake potential only for those types.
In the following, we default to \texttt{binary32}
because it is much faster than \texttt{binary64}.

\begin{figure}[bt]
\centering
\includegraphics[width=\linewidth]{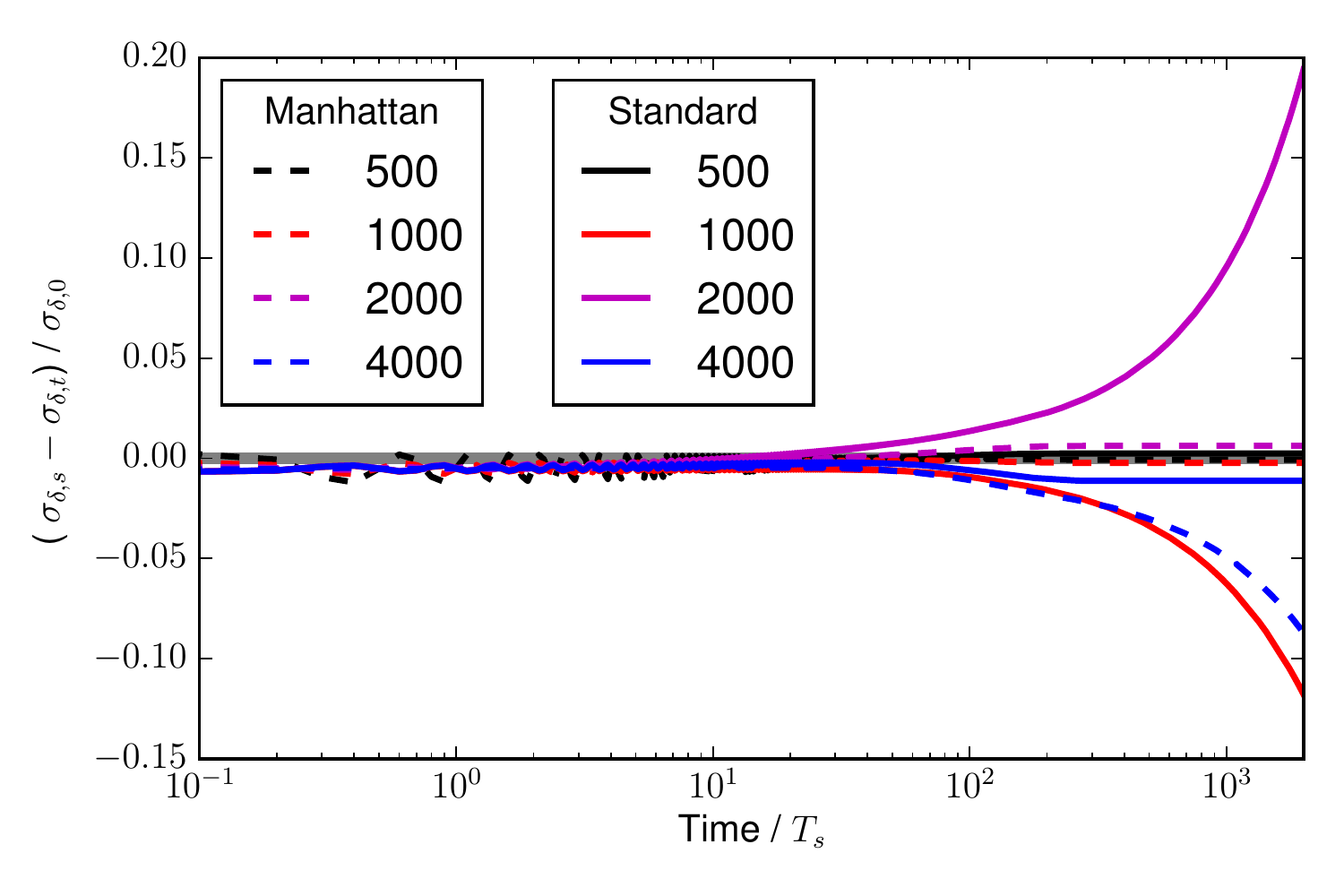}
\caption{
Exponential damping of the RMS energy spread for different numbers of
computation steps per synchrotron period.
(Here the grid size is set to 256.)
To emphasize the differences, the analytical, exponential
damping (\(\sigma_{\delta,t}\)) has been subtracted from the simulated values.
Every color represents a different number of
simulation steps per synchrotron period.
Until \(T=\unit[100]{T_s}\),
all simulations reproduce the exponential behavior.
Manhattan rotation (dashed lines) reproduces the set values
(solid gray line)
better than standard rotation (other solid lines)
independently of the number of steps.
However, when the number of steps per synchrotron period becomes too large,
the tiny step size leads to systematic drifts also
for this more robust method.
}
\label{fig: rotcompare}
\end{figure}

To investigate the influence of the size of the time steps,
we observe the evolution of a Gaussian distributed charge density
with \(\sigma_q = \sigma_p = 1.45\).
For this we set the damping time to \(\tau=45\,T_s\).
The Number of simulation steps (\(\Delta T\)) per \(T_s\) is
varied between 500 and 4000.
All settings reproduce the exponential damping.
However when \(\sigma_q\) approaches \(1\) some of the runs start to diverge.
As depicted in Fig.~\ref{fig: rotcompare} for
Manhattan rotation the error on the reconstructed values
is significantly lower (usually \(\ll 1\%\)).
Furthermore the Manhattan rotation is more robust against
changing the step sizes.
Note that the error increases when the number
of steps per synchrotron period becomes too large.
For standard rotation on the other hand there is no obvious
optimum for the size of the time steps.

\begin{figure}[bt]
\centering
\includegraphics[width=\linewidth]{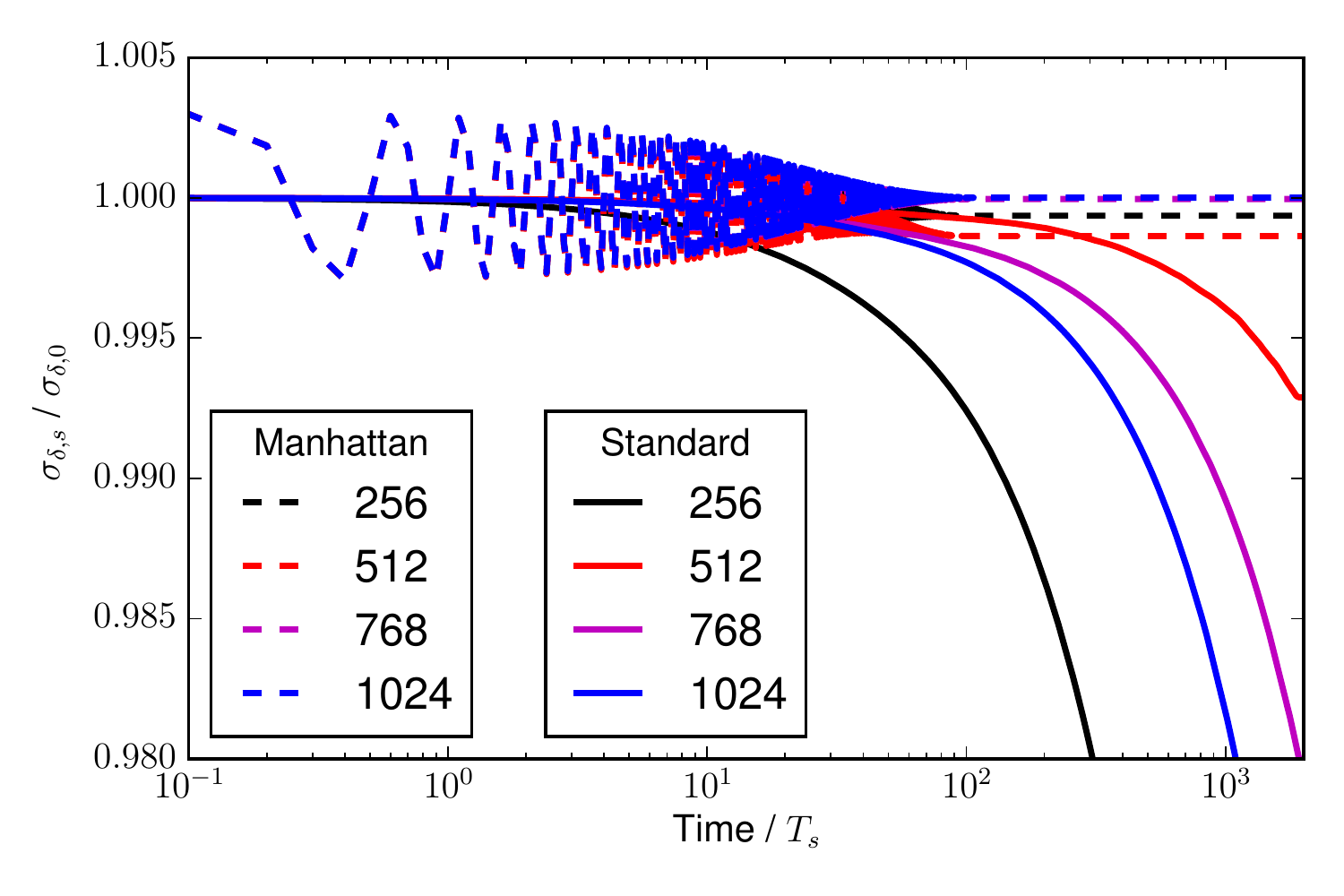}
\caption{
Evolution of the RMS energy spread for different grid sizes
and \(\Delta t = \unit[1/1000]{T_s}\).
Manhattan rotation (dashed lines) shows some initial wiggles
but reproduces the expected value
(\(\sigma_{\delta,t}=\sigma_{\delta,0}\), solid grey line)
very well.
For standard rotation (other solid lines) on the other hand
the divergence is even worse than shown in Fig.~\ref{fig: rotcompare}.
}
\label{fig: meshcompare}
\end{figure}

In another example to test the influence of the grid size
we study the evolution
of a Gaussian distributed charge density
with \(\sigma_q = \sigma_p = 1\).
From the physics point of view it is expected
that the distribution stays constant with time.
However, as illustrated in Fig.~\ref{fig: meshcompare},
it is observed that \(\sigma_p\) converges to different values or even diverges
depending on the numerical parameters.
For the standard rotation with the specific
time step of \(\Delta t = \unit[1/1000]{T_s}\),
there is no mesh size where the simulation converges.
Manhattan rotation on the other hand shows initial jitter with
a relative amplitude smaller than \(1\%\) and reliably converges afterwards.

\begin{table}[bt]
\caption{Typical relative errors observed in the convergence studies
using Manhattan rotation.}
\begin{center}
\begin{tabular}{l | r}
Parameter	& Error\\
\hline
Switching data types & \(< 0.1\%\)\\
Changing number of time steps	& \(<1\%\)\\
Changing number of grid points	& \(<1\%\)
\end{tabular}
\label{tab: maximum errors}
\end{center}
\end{table}

It can be seen that different combinations of the relative time step
\(\theta\), the damping factor \(\beta\), and the grid size may converge
to slightly different values -- or even diverge.
We found that Manhattan rotation is much more robust than the standard approach.
Provided that \(\theta\) and \(\beta\) do not become too small,
the relative error can be reliably kept below \(1\%\)
(see Tab.~\ref{tab: maximum errors}).

\subsection{Numerical artifacts}
\label{ssect: numerical artifacts}

Besides the numerical inaccuracies discussed above there are also
numerical artifacts. For those we identify two main sources:
interpolation and numerical differentiation.

\begin{figure}[tb]
\centering
\includegraphics[width=\linewidth]{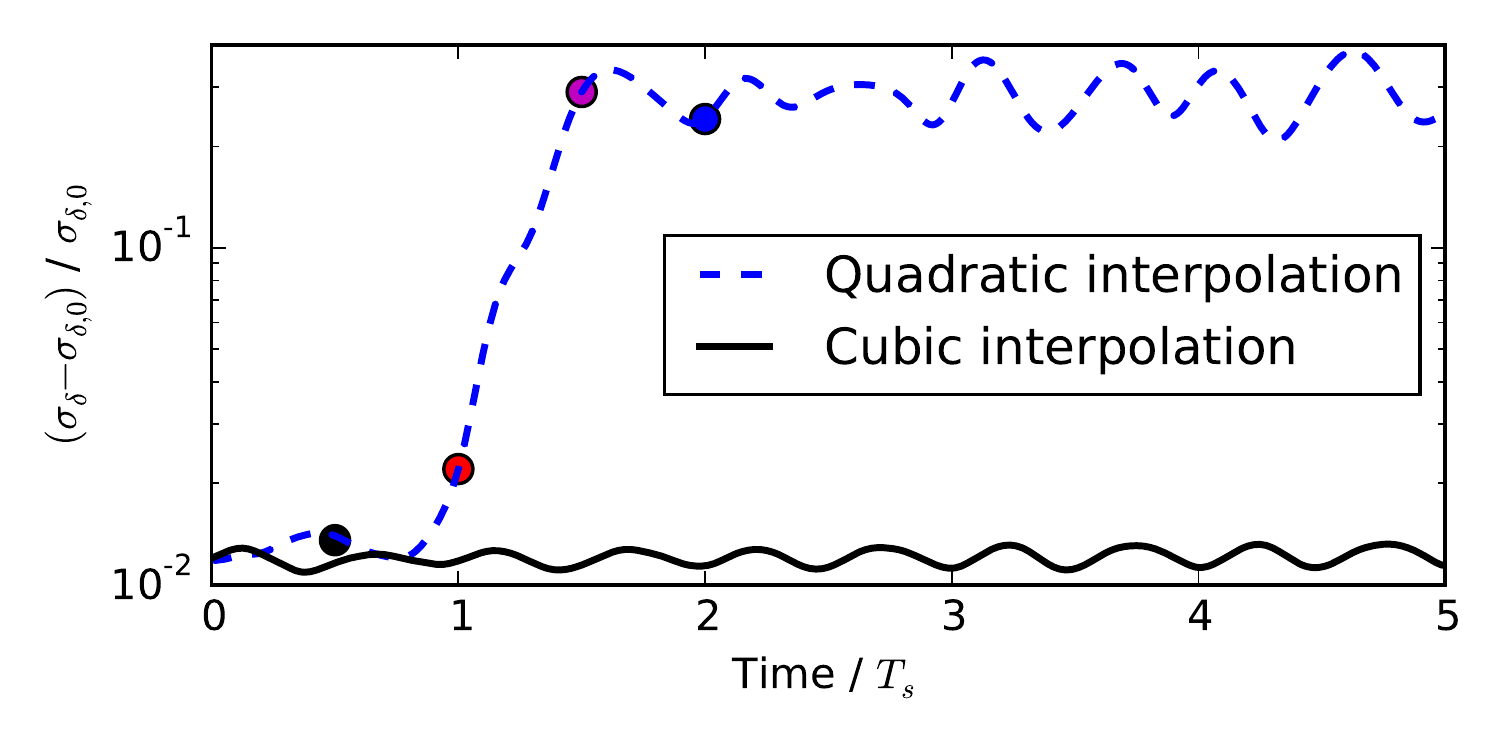}
\caption{
Evolution of the energy spread over time.
For the case where cubic interpolation is used (solid black line),
the energy spread stays at \(\sigma_\delta \approx 1.01 \sigma_{\delta,0}\).
When using quadratic interpolation (dashed blue line),
an increase in energy spread can be observed at \(T=1\;T_s\).
Looking at the bunch profiles at the points in time
marked by the disks (see Fig.~\ref{fig: interpolation-problems})
reveals that this increase is a numerical artifact.
}
\label{fig: interpolation-difference}
\end{figure}

As an example, we do two simulation runs with the same current distribution,
one run using cubic interpolation, the second using quadratic interpolation.
For the simulation run that uses cubic interpolation, the distribution
stays in a relatively calm state with just little oscillation.
Note that this is not equilibrium:
As expected for the unshielded CSR case and \(\xi>0.5\), we have
\(\sigma_\delta > \sigma_{\delta,0}\)~\cite{Stupakov2002}
and an oscillation with \(f \approx 2 f_s\).
The complete set of simulation parameters is listed
in Table~\ref{tab: artifact parameters}.
\begin{table}[b!]
\caption{Parameters for an example run to check for numerical artifacts.
For the given set no artifacts were observed.
Changing to quadratic interpolation however, triggered the
occurrence of (non-physical) structures with a period length of
two grid cells (see Fig.~\ref{fig: interpolation-problems}).
}
\begin{center}
\begin{tabular}{l c}
Parameter	& Value\\
\hline
Grid points per axis	& 256\\
Steps per \(T_s\)		& 4000 \\
Interpolation method	& cubic\\
Impedance model			& free space\\
Scaled current (\(\xi\))	& 0.516\\
Damping time			& \(200\;T_s\)
\end{tabular}
\label{tab: artifact parameters}
\end{center}
\end{table}
If there was no influence of the different interpolation schemes,
one would expect no difference between the two runs.
However, as Fig.~\ref{fig: interpolation-difference} depicts,
this is not guaranteed.

\begin{figure}[tb]
\centering
\includegraphics[width=\linewidth]{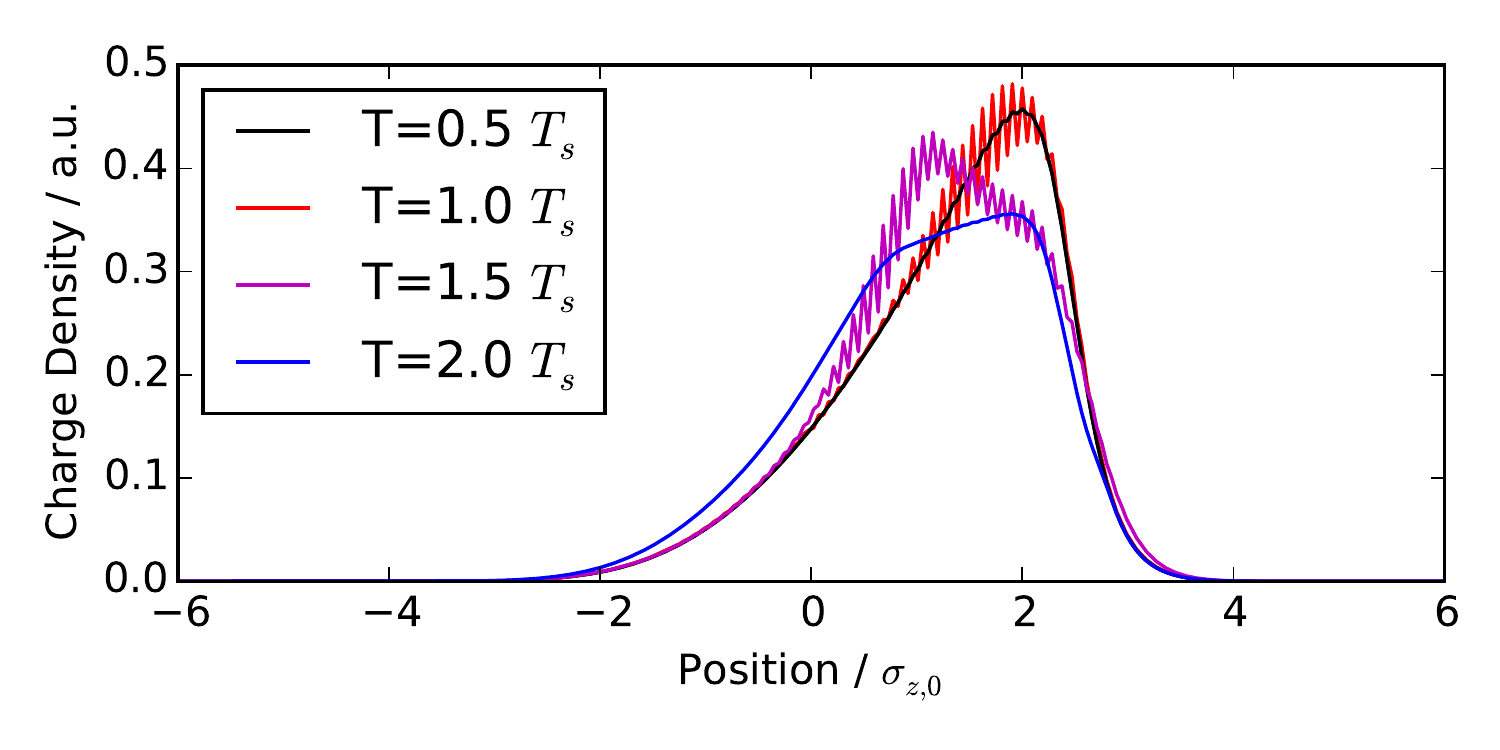}
\caption{
Bunch profiles of the simulation run using quadratic interpolation
at selected points in time
(cf. discs in Fig.~\ref{fig: interpolation-difference}).
The bunch profiles computed during the initial increase of the energy spread
show large ripples with a period length of two grid cells.
The earlier and later profiles do not show such structures.
This implies that the increase of energy spread
is driven by a numerical instability.}
\label{fig: interpolation-problems}
\end{figure}

The energy spread simulated using quadratic interpolation rapidly increases
to a higher value after one synchrotron period (\(T=1\;T_s\)).
On a longer time scale it will damp down again,
and after that a new numerical instability might rise.
If the aim is to find out whether the simulated conditions
are above the micro-bunching threshold these artifacts might not matter
-- below the threshold the initial, numerical modulation should not be amplified.
However, we want to track the evolution of the charge distribution,
and this numerical artifact might be interpreted as an unphysical
slow bursting frequency.
So we have to avoid numerical artifacts
as they occur in the run using quadratic interpolation
(depicted in Fig.~\ref{fig: interpolation-problems}):
The period length of the ripples is exactly two grid cells,
and they continue to exist even to a position where they
create negative charge densities.
In this particular case, the instability is clearly triggered
by numerical artifacts (overshoots) of the interpolation.

Note that also higher order polynomials show overshoots
and that also numerical differentiation can be expressed
in terms of interpolation polynomials.
In our tests, however, we did not find any case where a more complex
differentiation method than the one described in Eq.~\ref{eq: cubic diff}
was needed to avoid artifacts.
In contrast to the differentiation, there were rare cases where we
observed interpolation artifacts
even when using cubic interpolation polynomials.
Those we had to suppress by clamping (Eq.~\ref{eq: clamping})
-- which restricts interpolation to the range of the neighboring values.

\subsection{Comparison with measurements}
\label{ssec: measurements}

For our comparison with measurements results, we scale the
quantities that are a function of the revolution frequency
by a factor of \(2\pi R/C\) where \(R\) is the bending radius and
\(C\) is the circumference of ANKA.
This way measurements are comparable to the simulations -- which
assume an isomagnetic ring with the same bending radius.
The parameters used here are listed in Table~\ref{tab: iso-anka parameters}.
Note that this is just one set of possible parameters because
the magnet optics (and thus \(f_s\)) as well as the RF voltage \(V_{RF}\)
can be gradually changed at ANKA.

\begin{table}[b!]
\caption{Parameters of an isomagnetic accelerator comparable to ANKA
(\(t_d, h, f_s\), and \(f_{rev}\) are scaled by \(2\pi R/C=0.316\)).}
\begin{center}
\begin{tabular}{l c r}
Parameter	&	Symbol	& Value\\
\hline
Beam energy 			& \(E_0\)	& \(\unit[1.285]{GeV}\)\\
Energy spread			& \(\sigma_{\delta,0}\)	& \(0.47\times10^{-3}\)\\
Damping time			& \(t_d\)	& \(\unit[3.353]{ms}\)\\
Harmonic number			& \(h\)	& \(58.21\)\\
RF voltage				& \(V_{RF}\)	& \(\unit[1048]{kV}\)\\
Revolution frequency	& \(f_{rev}\)	& \(\unit[8.582]{MHz}\)\\
Synchrotron frequency	& \(f_s\)	& \(\unit[28.13]{kHz}\)\\
Vacuum chamber height	& \(g\)	& \(\unit[32]{mm}\)
\end{tabular}
\label{tab: iso-anka parameters}
\end{center}
\end{table}

\begin{figure*}[t]
\centering
\includegraphics[width=0.49\linewidth]{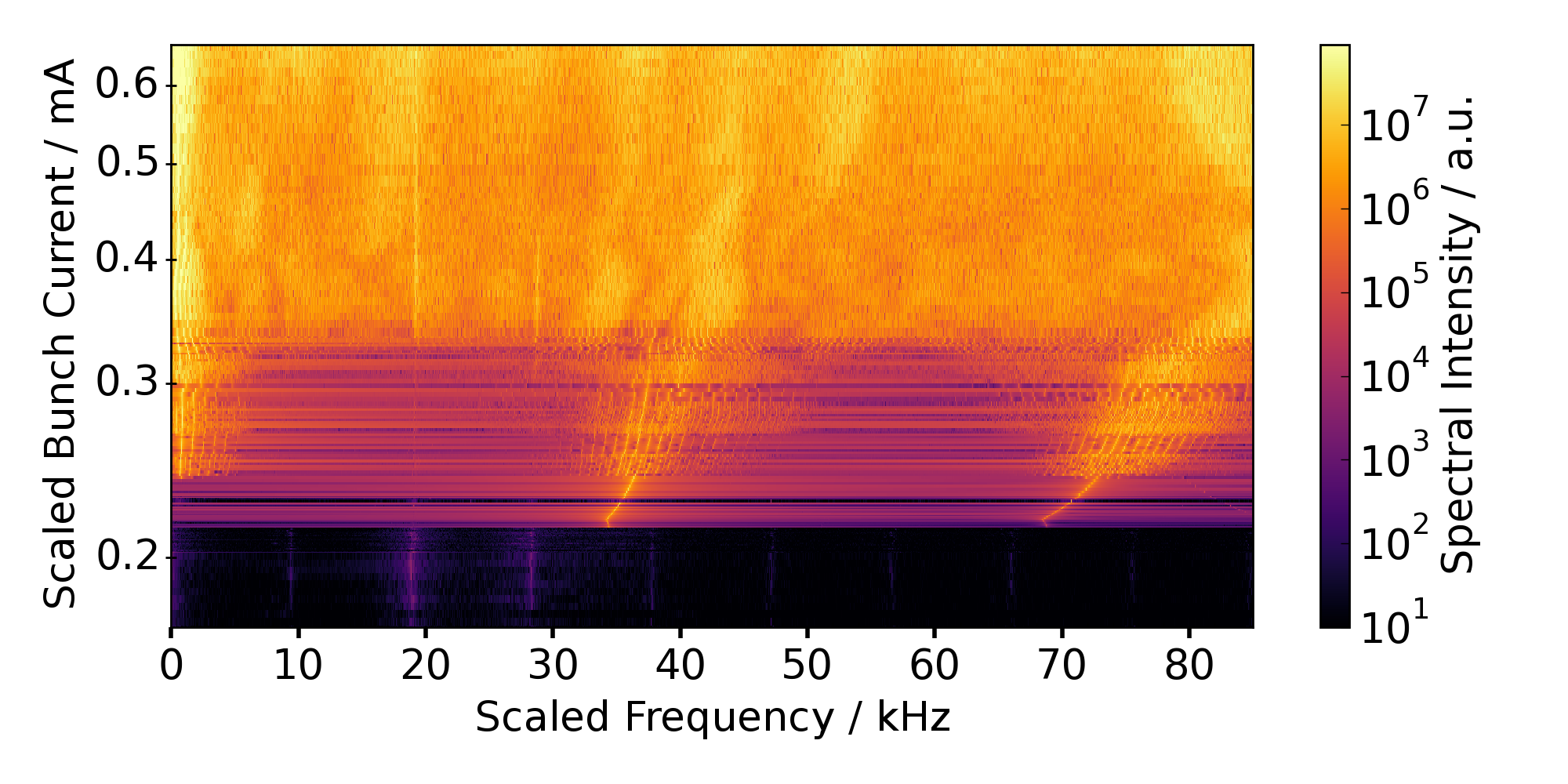}\hfill
\includegraphics[width=0.49\linewidth]{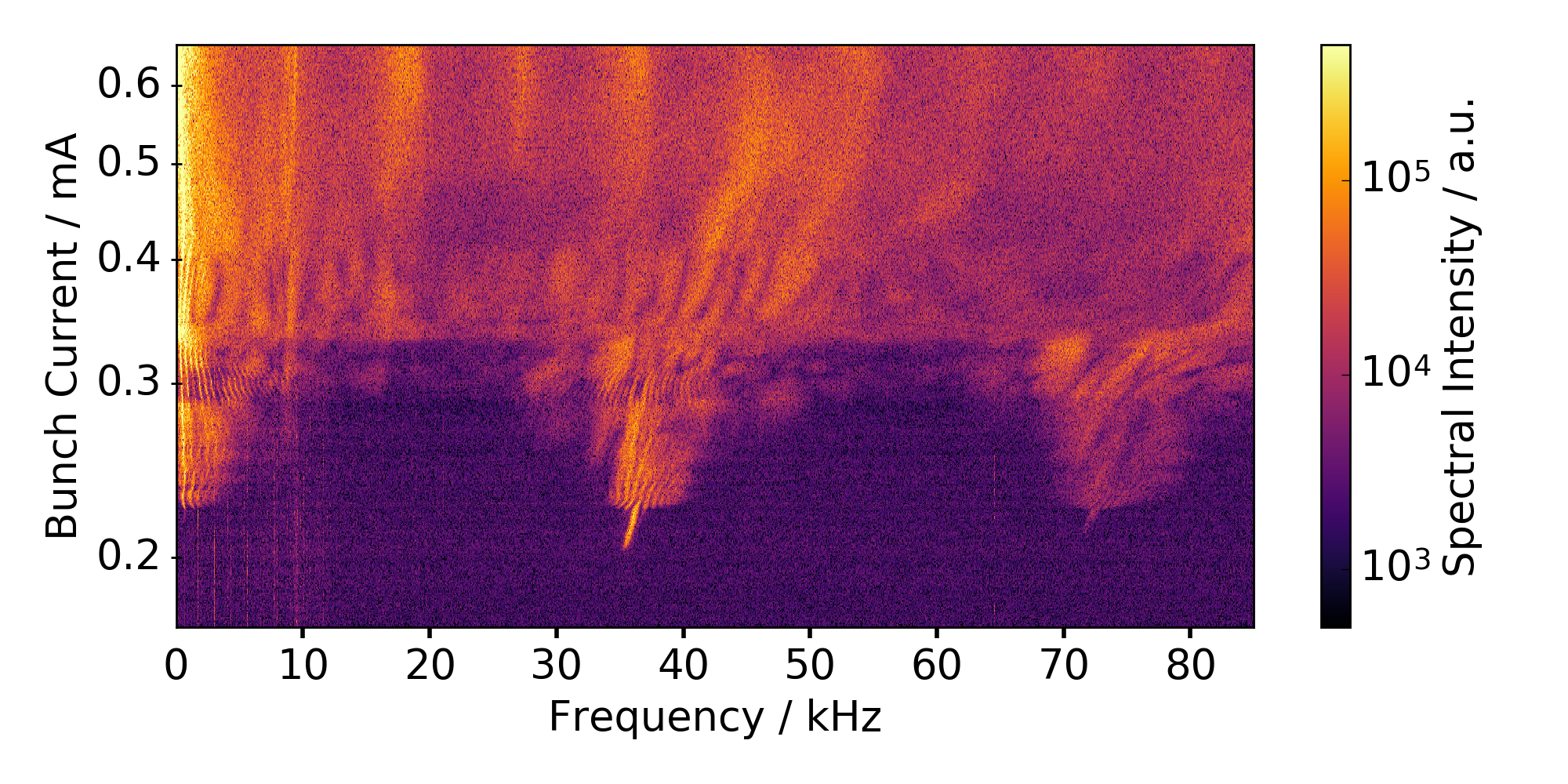}
\caption{
Example for a simulated (left) and a measured (right) bursting spectrogram.
For the simulated spectrogram the axis are scaled with a factor of
\(2\pi R/C\) to correct the mismatch due to the isomagnetic approximation.
There are small differences,
e.g. in the threshold current and in the frequencies.
However, keeping in mind the very simple model,
the general structure of the spectrograms matches quite well:
There is an isolated finger pointing down
(at \(f\approx\unit[35]{kHz}\));
the fingertip (\(I\approx\unit[0.21]{mA}\)) marks the instability threshold.
For slightly higher currents, fluctuations in the lower frequency range
(\(f < \unit[10]{kHz}\)) start and the finger broadens.
For the highest currents displayed here, there is a regime showing
parallel frequency lines that stay approximately constant with changing current.
}
\label{fig: spectrogram}
\end{figure*}

There are some effects we neglect for the simulation such as
the frequency response of the used detector,
and the coherent tune shift observed in the measurement.
Also, contributions from other impedances than shielded CSR
(e.g. geometric impedance) are not studied.

We do separate simulation runs for about 150 different currents between
\(I = \unit[1.3]{mA}\) and \(I = \unit[0.5]{mA}\).
For the first one, we start with the highest current
and a Gaussian charge distribution that is significantly
broader than the expected distribution.
To compensate for this,
we allow some extra time for convergence.
For the following simulation runs, we take the final
charge distribution of the run before that has the (slightly) higher current
as starting parameters.
(Different approaches to create starting distributions are discussed
in the Appendix~\ref{seq: starting distribution}.)
For each of these runs, the simulation time
on a AMD Radeon R9 290 graphics card is a bit more then ten minutes,
which makes a total simulation time of about 19 hours.

The spectrum of the emitted CSR is calculated using
\begin{equation}
	P(t,k) \propto \Re({Z_k}) \times |\tilde \rho(k,t)|^2,
\end{equation}
where \(\Re({Z_k})\) is the real part of the impedance and
\(|\tilde \rho(k,t)|^2\) the form factor of the bunch profile.
It is then integrated to obtain the power a detector would measure
\begin{equation}
	P(t) \propto \int P(t,k) \mathrm{d}k.
\end{equation}
In analogy to what is done for the measured data,
the resulting signal over time is Fourier transformed
to obtain a spectrogram of the `bursting' frequencies.
Figure~\ref{fig: spectrogram} shows this spectrogram of \(P(t)\).
The general structure
of the simulation and the measurement results agree very well:
The instability threshold is marked by the occurrence of an isolated finger
pointing down down to \(I\approx\unit[0.21]{mA}\)).
For slightly higher currents, the finger broadens and
fluctuations in the lower frequency range (\(f < \unit[10]{kHz}\)) start.
A third regime is observed at the highest currents.
It shows parallel frequency lines that stay approximately constant
with changing current.
There are slight mismatches,
e.g. in the threshold currents and in the frequencies,
but most features are well reproduced by the simulation.

This means that for ANKA not only the thresholds
(see also~\cite{2016arXiv160500536B}) but also 
the dynamics of the micro-bunching instability
are governed by an impedance that can be approximated
by the parallel plates CSR impedance.
To explain the details, a more complex model will be needed.
Two possibilities are to take into account higher orders
of the momentum compaction factor \(\alpha_c\)
or additional impedance contributions.

\section{Summary}

We introduced \emph{Inovesa}, a Vlasov-Fokker-Planck-solver
that uses a runtime-optimized implementation of the 
computation steps.
Utilizing OpenCL for parallelized computation,
it can simulate the dynamics of the longitudinal
phase space more than a 150 times faster
than a non-optimized implementation --
using a dedicated (consumer-grade) graphics card.
Furthermore, we eliminated sources of numerical artifacts and
have done numerical stability studies to show that
relative errors can usually be kept clearly below 1\%.

Using \emph{Inovesa} we were able to simulate the dynamics
of the longitudinal phase space of ANKA in the regime of the
micro-bunching instability.
To do so, we used an impedance model
that assumes CSR of electrons moving on a circular path
shielded by parallel plates.
Considering the simplicity of the model,
the numerical results show an excellent agreement
to the measurements.

\begin{acknowledgments}
C. Evain is acknowledged for useful discussions,
G. Stupakov for his suggestion to
investigate numerical effects of different rotation schemes.
We also thank M. Klein for giving us her code as a reference,
and T. Boltz for beta testing.
M. Brosi, P. Schönfeldt, and J. L. Steinmann
want to acknowledge the support by the
Helmholtz International Research School for Teratronics (HIRST).
\end{acknowledgments}

\appendix

\section{Starting Distribution}
\label{seq: starting distribution}

In equilibrium the energy is distributed according to a Gaussian function
and the bunch profile is described by the Ha\"issinski distribution~\cite{haissinski}.
However, here we are interested in the dynamics
of the micro-bunching instability
above the threshold current,
which means in non-equilibrium.
In this physical state,
there is no simple one-dimensional function
for any possible charge distribution.

One possibility is to start the simulation with a Gaussian distribution
that is broader than the expected charge distribution.
It will damp down until the physical state of the instability is reached.
Although for currents well above the instability threshold
any possible starting distribution will reach the same state,
we chose these initial conditions because it shows good convergence.
When alternatively using narrower distributions unphysical structures
form and might persist for a long time.

\begin{figure}[b]
\centering
\includegraphics[width=\linewidth]{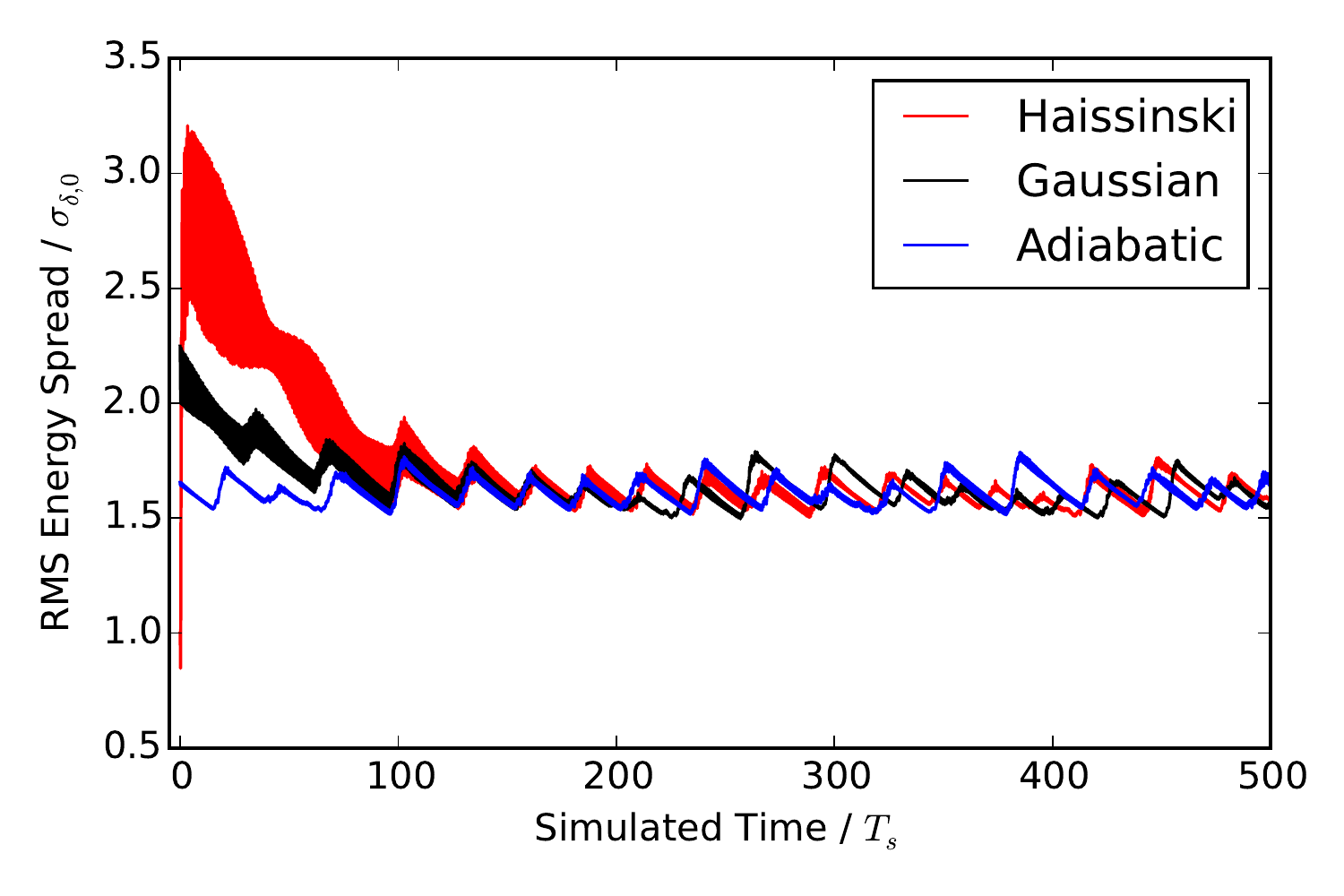}
\caption{
Evolution of the RMS energy spread over a time of 500 synchrotron periods
for \(I = \unit[1.5]{mA}\).
The width of the line is caused by high frequency oscillations.
For the starting distributions the Ha\"issinski distribution
(equilibrium at \(I = \unit[675]{\mu A}\)),
a Gaussian distribution with \(\sigma_p = \sigma_q = 2.25\),
and the final distribution of the previous run
(``adiabatic'', \(I = \unit[1.54]{mA}\)) were used.
Note that the Ha\"issinski distribution is immediately blown up and becomes
larger than the Gaussian distribution,
and that the high frequency oscillation is systematically
higher until \(T\approx \unit[300]{T_s}\).
}
\label{fig: anka-convergence}
\end{figure}

In Fig.~\ref{fig: anka-convergence} the evolution of the
RMS energy spread over 500 synchrotron periods is shown for three
different starting distributions at \(I=\unit[1.5]{mA}\).
When using the final distribution of a previous run
with slightly higher current
(here \(I = \unit[1.54]{mA}\)),
the simulation converges quasi instantaneously.
As shown, the Ha\"issinski distribution is immediately blown up and becomes
larger than the Gaussian distribution
that has been set to be larger than the expected distribution.
Also in the beginning (\(t < \unit[300]{T_s}\))
the oscillation is systematically enlarged.

\section{Running Inovesa}

\emph{Inovesa} is implemented as a non-interactive command-line tool.
To automatically simulate all data for a spectrogram as
shown in Fig.~\ref{fig: spectrogram},
a script may be used.
Using the ``adiabatic'' method discussed
in Appendix~\ref{seq: starting distribution}, an example bash script reads:
\begin{lstlisting}[language=Bash,breaklines=true,numbers=left,breakatwhitespace=true]
config="inovesa-run123.cfg"
lasti="500"
./inovesa -I ${last}e-6 --config $config -T 1500 -o $lasti.h5
for curri in {500..100..5}
do
  ./inovesa -I ${curri}e-6 -i $lasti.h5 -c $config -T 500 -o $curri.h5
  lasti=$curri
done
\end{lstlisting}

The parameters used here are:
\begin{description}
 \item[\texttt{--BunchCurrent}] (or \texttt{-I}) for the ring current
 due to a single bunch given (in Ampere)
 \item[\texttt{--config}] (or \texttt{-c}) the file name
 of a configuration file
 \item[\texttt{--rotations}] (or \texttt{-T}) the total simulation time
 (in synchrotron period lengths \(T_s\)))
 \item[\texttt{--InitialDistFile}] (or \texttt{-i}) the file name
 of an \emph{Inovesa} result file to use for the initial particle distribution
 \item[\texttt{--output}] (or \texttt{-o}) the file name to save results to
\end{description}

The configuration file used for the script contains all relevant parameters
in a key-value-representation.
Comments may be added using ``\texttt{\#}''.
Here is an example configuration file:
\begin{lstlisting}
BeamEnergy=1.3e+09 # in eV
BeamEnergySpread=0.00047 # relative
BendingRadius=5.559 # in m
BunchCurrent=1.2e-05 # in mA
DampingTime=0.01 # in ms
GridSize=256 # grid points per axis
HarmonicNumber=184 # f_RF/f_rev
PhaseSpaceSize=12 # sigma_z/E per axis
AcceleratingVoltage=1.4e+06 # in V
RevolutionFrequency=2.7e+06 # in Hz
alpha0=2e-4 # momentum compaction factor
VacuumGap=0.032 # full distance in m
gui=true # show live preview of results?
outstep=50 # write output every N steps
padding=8 # factor for zero-pading (FFT)
\end{lstlisting}
All parameters are optional:
If a parameter is not set \emph{Inovesa} will fall back to default values.
You might also overwrite settings from a configuration file
by passing the same parameter as a command line argument.
For short tests it is a good idea to enable the live preview
(\texttt{-g true}) and not to save the results
(by seting \texttt{-o /dev/null}).

\addcontentsline{toc}{section}{References}
\bibliographystyle{apsrev4-1}
\bibliography{inovesa}{}

\end{document}